\documentclass[12pt]{article}
\usepackage{latexsym,amssymb,amsmath}
\usepackage{amsthm, amstext}
\usepackage{array, amsfonts, mathrsfs}
\usepackage[dvips]{graphicx}

\usepackage{epsf}

\numberwithin{equation}{section}

\begin{document}
\date{}
\author{M.I.Belishev\thanks {Saint-Petersburg Department of
                 the Steklov Mathematical Institute, Saint-Petersburg State University, Russia;
                 belishev@pdmi.ras.ru.
                 }\,\, and A.F.Vakulenko\thanks {Saint-Petersburg Department of
                 the Steklov Mathematical Institute, Saint-Petersburg State University, Russia;
                 vak@pdmi.ras.ru.
                 }\,\,}

\title{Non-smooth unobservable states in control problem
for the wave equation in ${\mathbb R}^3$} \maketitle

\begin{abstract}
The paper deals with a dynamical system
\begin{align*}
&u_{tt}-\Delta u=0, \qquad (x,t) \in {\mathbb R}^3 \times (-\infty,0) \\
&u \mid_{|x|<-t} =0 , \qquad t<0\\
&\lim_{s \to \infty} su((s+\tau)\omega,-s)=f(\tau,\omega), \qquad
(\tau,\omega)   \in [0,\infty)\times S^2\,,
\end{align*}
where $u=u^f(x,t)$ is a solution ({\it wave}), $f \in {\cal F}
:=L_2\left([0,\infty);L_2\left(S^2\right)\right)$ is a {\it
control}. For the reachable sets ${\cal U}^\xi:=\{u^f(\cdot,
-\xi)\,|\,\, f \in {\cal F}\}\,\,(\xi\geqslant 0)$, the embedding
${\cal U}^\xi \subset {\cal H}^\xi:=\{y \in L_2({\mathbb
R}^3)\,|\,\,\,y|_{|x|<\xi}=0\}$ holds, whereas the subspaces
${\cal D}^\xi:={\cal H}^\xi \ominus {\cal U}^\xi$ of unreachable
({\it unobservable}) states are nonzero for $\xi> 0$. There was a
conjecture motivated by some geometrical optics arguments that the
elements of ${\cal D}^\xi$ are $C^\infty$-smooth with respect to
$|x|$. We provide rather unexpected counterexamples of $h\in {\cal
D}^\xi$ with ${\rm sing\,supp\,}h \subset \{x\in{\mathbb
R}^3|\,\,|x|=\xi_0>\xi\}$.
\end{abstract}

\section{Introduction}

\subsubsection*{Local control problem}
A dynamical system, which we deal with, is
\begin{align}
\label{Eq1} &u_{tt}-\Delta u=0, \qquad (x,t) \in {\mathbb R}^3 \times (-\infty,0) \\
\label{Eq2} &u \mid_{|x|<-t} =0 , \qquad t<0\\
\label{Eq3} &\lim_{s \to \infty}
su((s+\tau)\omega,-s)=f(\tau,\omega), \qquad (\tau,\omega)   \in
[0,\infty)\times S^2\,,
\end{align}
where $u=u^f(x,t)$ is a solution ({\it wave}), $f \in {\cal F}
:=L_2\left([0,\infty);L_2\left(S^2\right)\right)$ is a {\it
control}. The system describes propagation of incoming spherical
waves initiated by "infinitely far sources" (controls). The
relevant control problem is: given $y\in {\cal H} :=L_2({\mathbb
R}^3)$ to find $f \in \cal F$ obeying $u^f(\cdot,0)=y$. With
regard to hyperbolicity of the system, this problem can be
specified as follows.

Let ${\cal F}^\xi:=\{f \in {\cal F}\,|\,\,{\rm supp\,}f \subset
[\xi, \infty)\}$\,\,\,($\xi \geqslant 0$) be the subspace of
delayed controls ($\xi$ is the value of delay, ${\cal F}^0\equiv
\cal F$). With the system one associates a family of {\it
reachable sets} \footnote{below, the equality (*) follows from the
steady state property of the system: the operator $\Delta$
governing its evolution does not depend on time. By this, delay of
control leads to the same delay of wave.}
 $$
{\cal U}^\xi:=\{u^f(\cdot, 0)\,|\,\,f \in {\cal
F}^\xi\}\overset{*}= \{u^f(\cdot, -\xi)\,|\,\,f \in {\cal F}\}\,,
\quad \xi \geqslant 0.
 $$
Denote ${\cal H}^\xi:=\{y \in L_2({\mathbb
R}^3)\,|\,\,y\big|_{|x|<\xi}=0\}$, ${\cal H}^0\equiv\cal H$. The
waves, which are governed by the wave equation (\ref{Eq3}),
propagate with velocity $1$. At the moment $t=0$, the incoming
waves produced by delayed controls $f \in {\cal F}^\xi$ fill up
the part $\Omega^\xi:=\{x \in {\mathbb R}^3\,|\,\,|x|>\xi\}$ of
the space (see (\ref{Eq2})). Therefore, the embedding
 \begin{equation}\label{U xi subsetH xi}
{\cal U}^\xi \subset {\cal H}^\xi, \qquad \xi\geqslant 0
 \end{equation}
holds. The latter motivates the following setup of the {\it local
control problem} (LCP): given $y \in \cal H^\xi$ to find $f \in
\cal F^\xi$ obeying $u^f(\cdot,0)=y$. To study its solvability is
to clarify the character of the embedding (\ref{U xi subsetH xi}).
In particular, the question is whether the delayed waves
constitute an $L_2$-complete system in ${\cal H}^\xi$.

\subsubsection*{Results} For $\xi=0$, the Lax-Phillips theory
\cite{LP} provides ${\cal U}^0=\cal H$\, (see also \cite{BV2}). In
the mean time, as was first found out in \cite{BV2}, for $\xi>0$
the embedding (\ref{U xi subsetH xi}) is not dense:
 \begin{equation}\label{D^xi not= 0}
{\cal D}^\xi\,:=\,{\cal H}^\xi \ominus {\cal U}^\xi\,\not=\,\{0\},
\qquad \xi>0,
 \end{equation}
i.e., a lack of local controllability does occur. Elements of
${\cal D}^\xi$ are interpreted as {\it locally}
unreachable/unobservable states of system
(\ref{Eq1})--(\ref{Eq3}). Some of their properties are studied in
\cite{BV2}--\cite{BV5} \footnote{in particular, for the perturbed
equation $u_{tt}-\Delta u + qu=0$: see \cite{BV3}, \cite{BV5}}.

The subject of the paper is a smoothness of unobservable states.
In sec 3, we provide the arguments connected with propagation of
singularities (geometrical optics), which motivated our starting
conjecture: elements of ${\cal D}^\xi$ should be smooth at least
with respect to the radial variable $r=|x|$. Surprisingly, it
turns out to be wrong: we construct a counterexample of $h \in
{\cal D}^\xi$ such that ${\rm sing\, supp\,}h=\{x \in {\mathbb
R}^3\,|\,\,|x|=\xi_0>\xi\}$. It is this construction, which is our
main result.

\subsubsection*{Comments}
$\bullet$\,\,\,In sec 2, we discuss an LCP for a bounded domain (a
ball). The goal is to emphasize the principal difference: in a
ball, for the straightforward analog of system
(\ref{Eq1})--(\ref{Eq3}) one has ${\cal D}^\xi=\{0\}$, i.e. the
local controllability does hold \footnote{more precisely, we speak
about a local approximate boundary controllability
\cite{ABI}--\cite{BL}}. In the mean time, system
(\ref{Eq1})--(\ref{Eq3}) in the space (and the LCP for it) is a
relevant limit case of the system in a ball. So, a notable effect
occurs: propagating from infinity, the waves lose completeness in
the filled domains.
\smallskip

\noindent$\bullet$\,\,\,Local completeness of waves produced by
boundary controls plays a crucial role in the {\it boundary
control method}, which is an approach to inverse problems based on
their relations to control theory \cite{BIP}. Planning to apply
the BC-method to the dynamical inverse scattering problem for the
acoustical equation $u_{tt}-\Delta u + qu=0$ in ${\mathbb R}^3$,
we have to bear in mind (\ref{D^xi not= 0}) as a fact, which
complicates such an application. It is the reason, which motivates
our interest to unobservable states.
\smallskip

\noindent$\bullet$\,\,\,The work is supported by the grants RFBR
11-01-00407A and SPbGU\\ 11.38.63.2012, 6.38.670.2013.

\section{Problem in a ball}
\subsubsection*{System $\alpha^T$}
Denote $B^s:=\{x \in {\mathbb R}^3\,|\,\,|x|<s\},\,\,\Gamma^s:=\{x
\in {\mathbb R}^3\,|\,\,|x|=s\}$; fix a positive $T$. Here we deal
with a dynamical system $\alpha^T$ of the form
 \begin{align}
\label{Eq1T} &u_{tt}-\Delta u=0 && {\rm in}\,\,\,B^T \times (0,T) \\
\label{Eq2T} &u \mid_{t=0} = u_t\mid_{t=0}=0 && {\rm in}\,\,\,\overline{B^T}\\
\label{Eq3T} & u=f && {\rm on}\,\,\,[0,T] \times \Gamma^T\,,
 \end{align}
where $u=u^f(x,t)$ is a solution ({\it wave}), $f \in {\cal F}^T
:=L_2\left([0,T];L_2\left(\Gamma^T\right)\right)$ is a {\it
control}. The following is the standard control theory  attributes
of the system.

\noindent$\bullet$\,\,\,The {\it outer space} is ${\cal F}^T$. It
contains the subspaces $${\cal F}^{T, \xi}:=\{f \in {\cal
F}^T\,|\,\,{\rm supp\,}f \subset [\xi,T]\}, \qquad 0\leqslant \xi
\leqslant T$$ (${\cal F}^{T,0}\equiv {\cal F}^T$), which consist
of the delayed controls, $\xi$ being the delay, $T-\xi$ the action
time.

\noindent$\bullet$\,\,\,The {\it inner space} of states ${\cal
H}^T:=L_2(B^T)$ contains the subspaces
$${\cal H}^{T, \xi}:=\{y \in {\cal H}^T\,|\,\,{\rm supp\,}y \subset
\overline {\Omega^{T, \xi}}\}, \qquad 0\leqslant \xi \leqslant T$$
(${\cal H}^{T,0}\equiv {\cal H}^T$), where $\Omega^{T,\xi}:=\{x
\in B^T\,|\,\,\xi < |x|<T\}$.

\noindent$\bullet$\,\,\,The `input $\to$ state' correspondence is
realized by a {\it control operator} $W^T: {\cal F}^T \to {\cal
H}^T,\,\,\,W^Tf:=u^f(\cdot,T)$. This operator is bounded
\cite{LLT}.

\subsubsection*{Local controllability}
The waves governed by equation (\ref{Eq1T}) propagate in $B^T$
with velocity $1$. Delay of control implies the same delay of
wave. As a result, one has
 \begin{equation}\label{supp u}
{\rm supp\,}u^f(\cdot,T)\,\subset \,\overline{\Omega^{T,\xi}}
\qquad {\rm for}\,\,\,f \in {\cal F}^{T,\xi},
 \end{equation}
so that at the final moment the delayed waves fill up the
near-boundary layer of the thickness $T-\xi$.

Introduce the {\it reachable sets}
$${\cal U}^{T,\xi}:=\{u^f(\cdot,T)\,|\,f \in {\cal F}^{T,\xi}\}=W^T {\cal F}^{T,\xi}.$$
By (\ref{supp u}), the embedding ${\cal U}^{T,\xi}\subset {\cal
H}^{T,\xi}$ holds. The well-known fact is that this embedding is
dense:
 \begin{equation}\label{U xi=H xi}
\overline {{\cal U}^{T,\xi}}\,=\,{{\cal H}^{T,\xi}}\,, \qquad
0\leqslant \xi \leqslant T
 \end{equation}
(the closure in ${\cal H}^T$). It is derived from the fundamental
Holmgren-John uniqueness theorem (see, e.g., \cite{LP}, Chaper IV,
Theorem 1.5) by the scheme originated by D.Russell in
\cite{Russell}.

Thus, the waves do constitute complete systems in the filled
domains $\Omega^{T,\xi}$, i.e., system $\alpha^T$ is locally
controllable from the boundary. In addition, note that for $\xi<T$
one has ${{\cal H}^{T,\xi}}\backslash{{\cal
U}^{T,\xi}}\not=\emptyset$, so that the reachable set is dense in
the subspace ${\cal H}^{T,\xi}$ but does not exhaust it
\cite{ABI}. Therefore, one should speak about {\it approximate}
local boundary controllability.

\subsubsection*{System $\alpha^T_*$}
A dynamical system dual to $\alpha^T$ is
 \begin{align}
\label{Eq1T*} &v_{tt}-\Delta v=0 && {\rm in}\,\,\,B^T \times (0,T) \\
\label{Eq2T*} &v \mid_{t=T}=0,\,\,v_t\mid_{t=T}=y && {\rm in}\,\,\,\overline{B^T}\\
\label{Eq3T*} & v=0 && {\rm on}\,\,\, [0,T] \times \Gamma^T\,,
 \end{align}
where $y \in {\cal H}^T$ and $v=v^y(x,t)$ is a solution. With this
system one associates an {\it observation operator} $O^T: {\cal
H}^T \to {\cal F}^T$,
 $$
(O^T y)(t)\,:=\,v^y_r (\cdot,t)\big |_{\Gamma^T}, \qquad
0\leqslant t \leqslant T\,,
 $$
where $(\,\,)_r:=\frac{\partial}{\partial |x|}\,$. Integration by
parts leads to the relation
 \begin{equation}{\label{dual rel}}
O^T\,=\,(W^T)^*\,,
 \end{equation}
which motivates the term `dual'. Elements of ${\rm Ker\,}O^T$ are
said to be {\it unobservable}. Since
 $$
{\rm Ker\,}O^T={\cal H}^T \ominus \overline{{\rm
Ran\,}(O^T)^*}\overset{(\ref{dual rel})}=
 {\cal H}^T \ominus \overline{{\rm Ran\,}W^T}=
 {\cal H}^T \ominus \overline{{\cal U}^{T,T}}\overset{(\ref{U xi=H xi})}=\{0\},
 $$
there are no unobservable elements in the dual system.

\subsubsection*{Propagation of singularities in $\alpha^T_*$}
Now, fix a positive $\xi_0<T$. In (\ref{Eq2T*}), let us take $y
\in {\cal H}^T$ provided ${\rm supp\,}y \subset B^T$\, (so that
$y$ vanishes near $\Gamma^T=\partial B^T$), $y\in
C^\infty(B^T\backslash \Gamma^{\xi_0})$, and
 \begin{equation}\label{sing supp y (ball)}
\emptyset \not={\rm sing\, supp\,}y \subset \Gamma^{\xi_0}.
 \end{equation}
By standard propagation of singularity arguments (see, e.g.,
\cite{Horm}), the latter yields
 \begin{equation}\label{sing supp v^y (ball)}
\emptyset \not= {\rm sing\, supp\,}v^y\,\subset\,C^{T,\xi_0}_1
\cup C^{T,\xi_0}_2 \cup C^{T,\xi_0}_3\,,
 \end{equation}
where
 \begin{align*}
& C^{T,\xi_0}_1:=\{x \in
\overline{B^T}\,|\,\,|x|=-t+(T-\xi_0)\},\,\quad
C^{T,\xi_0}_2\,:=\,\\
&=\,\{x \in \overline{B^T}\,|\,\,|x|=|t-(T-\xi_0)|\}, \,\,\,\,
C^{T,\xi_0}_3:=\{x \in \overline{B^T}\,|\,\,|x|=t+(T-\xi_0)\},
 \end{align*}
are the characteristic cones (Fig 1a).
 \begin{figure}[h]
  \begin{center}
  \epsfysize=8cm
  \epsfbox{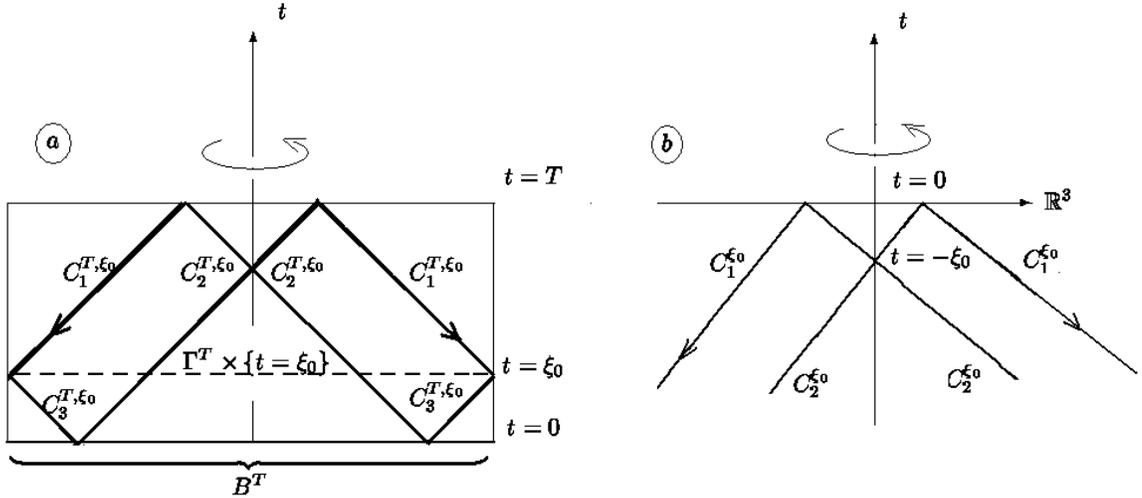}
  \end{center}
 \caption{Characteristic cones}
  \end{figure}
The cone $C^{T,\xi_0}_1$ supports the singularities of $v^y(\cdot,
t)$, which move (in the reversed time) towards the boundary
$\Gamma^T$ and reach it at the moment $t=\xi_0$. The singularities
reflected from the boundary are supported in $C^{T,\xi_0}_3$. As a
result, for the trace of the radial derivative $v^y_r$ at the
boundary (as a function on $\Gamma^T \times (0,T)$) one has
 \begin{equation*}
\emptyset \not={\rm sing\, supp\,} v^y_r \subset \Gamma^T \times
\{t=\xi_0\}\,.
 \end{equation*}
So, {\it any} `spherical' singularity (\ref{sing supp y (ball)})
manifests itself at the boundary at the proper moment $t=\xi_0$.
\smallskip

As example, consider $y$ with a radial jump. Denote
$r=|x|,\,\,\omega=\frac{x}{|x|}\in S^2$, so that $x=r\omega$ and
$\Gamma^{\xi_0}=\{x=\xi_0 \omega\,|\,\,\omega \in S^2\}$. Assume
that
 \begin{align*}
\lim \limits_{\varepsilon \to
0}\left[y\left((\xi_0+\varepsilon)\omega\right)-y\left((\xi_0-\varepsilon)\omega\right)\right]
\,=\,\alpha(\omega), \quad \omega \in S^2 .
 \end{align*}
with a nonzero $\alpha \in C^\infty(S^2)$. Then, $v^y_t$ and
$v^y_r$ have a jump supported on the cones $C^{T,\xi_0}_i$. The
jump at $C^{T,\xi_0}_1$ can be treated by standard geometrical
optics considerations, which provide
 \begin{align} \notag & \lim
\limits_{\varepsilon \to
0}\left[v^y_r\left((\xi_0+T-t+\varepsilon)\omega,t\right)-v^y_r
\left((\xi_0+T-t-\varepsilon)\omega,t\right)\right]\,\\
 \label{jump d/dr v^y (ball)}
 &=\,- \frac{\xi_0}{\xi_0+T-t} \alpha(\omega),
\qquad \omega \in S^2,\,\,\xi_0<t<T
 \end{align}
(see, e.g., \cite{Ikawa}, Chapter 3). As a result, the trace of
$v^y_r$ on the boundary turns out to be discontinuous at
$t=\xi_0$, and quite standard technique \cite{Ikawa} provides
 \begin{align*}
\lim \limits_{\varepsilon \to
0}\left[v^y_r(T\omega,\xi_0+\varepsilon)-v^y_r(T\omega,\xi_0-\varepsilon)\right]\,=\,-
\frac{\xi_0}{T} \alpha(\omega), \quad \omega \in S^2 .
 \end{align*}
The latter is equivalent to
 \begin{align}\label{jump O^Ty (ball)}
\lim \limits_{\varepsilon \to 0}\left[\left(O^T
y\right)(\xi_0+\varepsilon)-\left(O^Ty\right)(\xi_0-\varepsilon)\right]\big|_{x=T\omega}\,=\,-
\frac{\xi_0}{T} \alpha(\omega), \quad \omega \in S^2 .
 \end{align}
Recall that here $O^Ty \in {\cal F}^T$ is an
$L_2(\Gamma^T)$-valued function of $t \in [0,T]$, so that its jump
at $t=\xi_0$ is an element of $L_2(\Gamma^T)$, its value being
taken at the point $x=T\omega \in \Gamma^T$. This jump is
observable in any time interval $[\xi,T]$ as $\xi<\xi_0$.

\section{Problem in ${\mathbb R}^3$}

\subsubsection*{System $\alpha$}
Here we deal with the system $\alpha$ of the form
 \begin{align}
\label{Eq1 again} &u_{tt}-\Delta u=0, \qquad (x,t) \in {\mathbb R}^3 \times (-\infty,0) \\
\label{Eq2 again} &u \mid_{|x|<-t} =0 , \qquad t<0\\
\label{Eq3 again} &\lim_{s \to \infty}
su((s+\tau)\omega,-s)=f(\tau,\omega), \qquad (\tau,\omega)   \in
[0,\infty)\times S^2\,,
 \end{align}
where $u=u^f(x,t)$ is a solution ({\it wave}); $f$ is a {\it
control}, which we regard as an $L_2\left(S^2\right)$-valued
function of $\tau \in [0,\infty)$. For a smooth enough $f$
vanishing near $\tau=0$, problem (\ref{Eq1 again})--(\ref{Eq3
again}) has a unique classical solution, and the well-known
representation
 \begin{equation}\label{u via f}
u^f(x,t)\,=\,\frac{1}{2\pi}\int_{S^2}\tilde f_\tau(t+x\cdot
\omega, \omega)\,d\sigma_\omega
 \end{equation}
holds, where $\tilde f$ is extension of $f$ from $[0,\infty)$ to
$(-\infty,\infty)$ {\it by zero},
$(\,\,)_\tau:=\frac{\partial}{\partial \tau}$, $"\cdot"$ is the
standard inner product in ${\mathbb R}^3$, and $d\sigma$ is the
surface element on $S^2$ \,(see, e.g., \cite{LP}).
\smallskip

The following is the control theory attributes of system $\alpha$.

\noindent$\bullet$\,\,\,The {\it outer space} is ${\cal F}
:=L_2\left([0,\infty);L_2\left(S^2\right)\right)$. It contains the
subspaces
 $${\cal F}^\xi := \{f \in {\cal F}\,|\,\,{\rm supp\,}f
\subset [\xi,\infty)\}, \qquad \xi \geqslant 0
 $$ (${\cal
F}^{0}\equiv \cal F$). The elements of ${\cal F}^\xi$ are
interpreted as delayed controls, $\xi$ being the value of delay.

\noindent$\bullet$\,\,\,The {\it inner space} of states ${\cal
H}:=L_2({\mathbb R}^3)$ contains the subspaces
 $$
{\cal H}^\xi=\{y \in {\cal H}\,|\,\,{\rm supp\,}y \subset
\overline {\Omega^\xi}\}, \qquad \xi \geqslant 0
 $$ (${\cal H}^{0}\equiv \cal
H$), where $\Omega^\xi:=\{x \in {\mathbb R}^3\,|\,\,|x|>\xi\}$.

\noindent$\bullet$\,\,\,The `input $\to$ state' correspondence is
realized by a {\it control operator} $W: {\cal F} \to {\cal
H},\,\,\,W f:=u^f(\cdot,0)$. By (\ref{u via f}), for available
controls one has
 \begin{equation}\label{repres W}
(W f)(x)\,=\,\frac{1}{2\pi}\int_{S^2}\tilde f_\tau(x \cdot \omega,
\omega)\,d\sigma_\omega\,.
 \end{equation}
Moreover, $W$ is a unitary operator, \cite{BV2}, \cite{BV3}), so
that $W^*\,=\,W^{-1}$ holds  (see, e.g., \cite{LP}.
\medskip

\subsubsection*{Controllability. Subspaces ${\cal D}^\xi$.}
System $\alpha$ can be regarded as a limit case of system
$\alpha^T$ \footnote{Namely, let $f \in {\cal F}$ and
$f_T:=f|_{0\leqslant t \leqslant T}\in {\cal F}^T$, $u^{f_T}$ the
solution to (\ref{Eq1T})--(\ref{Eq3T}). Then quite simple
considerations provide
 \begin{equation*}
u^f(x,t)\,=\,\lim \limits_{T\to \infty}T u^{f_T}(x,t+T), \qquad
(x,t) \in {\mathbb R}^3 \times (-\infty,0)\,.
 \end{equation*}}.
However, they substantially differ in controllability properties.

In system $\alpha$, a relevant analog of the reachable sets ${\cal
U}^{T,\xi}$ of $\alpha^T$ is
 $$
{\cal U}^\xi\,:=\,W{\cal F}^\xi\,\overset{**}=\,\{u^f(\cdot,
-\xi)\,|\,\,f \in {\cal F}\}, \qquad \xi \geqslant 0\,.
 $$
Since $W$ is a unitary operator, ${\cal U}^\xi$ is a closed
subspace \footnote{in contrast to ${\cal U}^{T,\xi}$ for $\xi<T$
in a ball}. The equality $(**)$ is easily seen from (\ref{Eq2
again}), (\ref{u via f}) and corresponds to the steady-state
property. Also, for a delayed $f \in {\cal F}^\xi$, (\ref{repres
W}) evidently implies $u^f(\cdot,0)|_{|x|<\xi}=0$, i.e., $ {\rm
supp\,}u^f(\cdot,0)\subset \overline{\Omega^\xi}\,.$ Thus, at the
final moment $t=0$, the delayed waves incoming from infinity fill
up the domain $\Omega^\xi$, and the embedding ${\cal U}^\xi
\subset {\cal H}^\xi$ holds. The question is whether they are
complete in ${\cal H}^\xi$, i.e., the local controllability
occurs. The answer is negative: as was found out in \cite{BV2},
\cite{BV3}, in contrast to (\ref{U xi=H xi}), one has
 \begin{equation*}
{\cal D}^\xi\,:=\,{\cal H}^\xi \ominus {\cal U}^\xi\,\not=\,\{0\},
\qquad \xi>0\,.
 \end{equation*}
In the mean time, for $\xi=0$ the equality ${\cal U}^0={\cal
H}^0=\cal H$ does hold.
\smallskip

Elements of ${\cal D}^\xi$ are said to be the states {\it
unreachable for controls} $f \in {\cal F}^\xi$. Later on, they
will be also specified as {\it unobservable} in a proper sense.
Their properties is the main subject of our paper. Recall some of
the known facts.

For an integer $l\geqslant 1$, denote
$\sigma(l):=\max\{j=0,1,2,\dots\,|\,\,l-2j>0\}$. Introduce a class
of polynomials of variable $s>0$
 $$
{\cal P}_l\,:=\,{\rm span\,}\{s^{l-2j}\}_{j=0}^{\sigma(l)}\,.
 $$
Let $Y_l^m$ be the standard spherical harmonics on $S^2$:
 $$
-\Delta_\omega
Y_l^m=l(l+1)Y_l^m,\,\,\,\,(Y_l^m,Y_{l'}^{m'})_{L_2(S^2)}=
\begin{cases}
1 &{\rm if}\,\,l=l',m=m'\\
0 &{\rm otherwise}
\end{cases}
 $$
where $Y_0^0 = {\rm const},\,\,m\in \{-l,-l+1,\dots,0, \dots,
l-1,l\}$, and $\Delta_\omega$ is the Beltrami-Laplace operator on
$S^2$. As is shown in \cite{BV2}, \cite{BV3}, the representation
 \begin{align}\label{D=sun Dl}
\notag &{\cal D}^\xi =\oplus \sum \limits_{l\geqslant 1}{\cal
D}^\xi_l\,,\qquad
{\cal D}^\xi_l\,:=\\
&=\,\left\{y \in {\cal H}^\xi \,\bigg|\,\,
y\big|_{|x|>\xi}=\frac{1}{r}\,
p_l\!\left(\frac{1}{r}\right)Y_l(\omega)\,\,p_l \in {\cal
P}_l,\,\,Y_l \in {\rm span\,}\{Y_l^m\}_{m =-l}^l\right\}
 \end{align}
is valid (recall that $x=r\omega, \,r=|x|,\,
\omega=\frac{x}{|x|}$). One more characterization is
 \begin{align}\label{polyharmonic}
{\cal D}^\xi_l =\,\left\{y \in {\cal H}^\xi\,\bigg|\,\,\Delta^l
y=0\,\,\,\,{\rm in}\,\, \Omega^\xi,\,\,y(r, \cdot)\big|_{r>\xi}
\in {\rm span\,}\{Y_l^m\}_{m =-l}^l \right\}\,,
 \end{align}
where $y(r,\omega):=y(r\omega)=y(x)$, so that ${\cal D}^\xi_l$
consists of $l$-polyharmonic functions. Note that ${\cal D}^\xi_l
\subset C^{\infty}(\Omega^\xi)$ holds.
\smallskip

\noindent{\bf Remark} A lack of local controllability in system
$\alpha$ is, in a sense, partially compensated by the following
property. It can be shown that for any $0 < \xi < \xi'<\infty$ one
has
 \begin{equation*}
\overline{\left \{ u^f(\cdot,0) \big|_{ B^{\xi'} \setminus B^\xi}
\,\bigg |\,\, f \in {\cal F}^\xi \right \}} =L_2(B^{\xi'}
\setminus B^\xi),
 \end{equation*}
i.e., the forward parts of delayed incoming waves possess a local
completeness.

\subsubsection*{System $\alpha_*$} The system dual to $\alpha$
is of the form
 \begin{align}
\label{Eq1*} &v_{tt}-\Delta v=0 && {\rm in}\,\,\,{\mathbb R}^3 \times (-\infty,0) \\
\label{Eq2*} &v \mid_{t=0}=0,\,\,v_t\mid_{t=0}=y && {\rm in}\,\,\,
{\mathbb R}^3
 \end{align}
where $y \in \cal H$ and $v=v^y(x,t)$ is a solution. For a smooth
$y$, by Kirchhoff, one has
 \begin{equation*}
v^y(x,t)\,=\,\frac{1}{4\pi t}\int_{\Gamma^{|t|}_x}
y(\gamma)\,d\sigma_\gamma, \qquad t<0\,,
 \end{equation*}
where $\Gamma^s_x:=\{\gamma \in {\mathbb
R}^3\,|\,\,|x-\gamma|=3\}$.

 With the dual system one associates an {\it observation
operator} $O:{\cal H} \to {\cal F}$ defined by
 \begin{equation}\label{def O in mathbb RR3^3}
(Oy)(\tau,\omega):= \lim_{s \to \infty} s  \left [v^y_t
 + v^y_r\right ]((s+\tau) \omega,-s), \,\,(\tau,\omega) \in [0,\infty)
\times S^2
 \end{equation}
on $C^\infty$-smooth rapidly decaying functions. The duality
relation is
 \begin{equation}\label{O=W^*= Radon}
O\,=\,W^*\,=\,-\frac{1}{4\pi}\frac{\partial}{\partial \tau}{\cal
R}\,,
 \end{equation}
where $({\cal
R}y)(\tau,\omega):=\int_{x\cdot\omega=\tau}y(x)\,ds_x$ is the
Radon transform: see \cite{LP}, \cite{BV3}, \cite{BV5}.
\smallskip

For any $h \in {\cal D}^\xi$ and $f \in {\cal F}^\xi$, one has
 \begin{align*}
 \int_\xi^\infty
\left((Oh)(\tau),f(\tau)\right)_{L_2(S^2)}\,d\tau=(Oh,f)_{\cal
F}\overset{(\ref{O=W^*= Radon})}=(h,Wf)_{\cal H}
=\left(h,u^f(\cdot,0)\right)_{\cal H}=0
 \end{align*}
since $u^f(\cdot,0)\in {\cal U}^\xi$. By arbitrariness of $f$, we
get $Oh\big|_{\tau\geqslant \xi}=0$, so that the elements of
${\cal D}^\xi$ can be specified as {\it unobservable} (at
infinity) in the interval $\xi \leqslant \tau<\infty$.

\subsubsection*{Propagation of jumps in $\alpha_*$}
Fix $0<\xi<\xi_0$. Let $y \in \cal H$ in (\ref{Eq2*}) be such that
${\rm supp\,}y \subset \overline{\Omega^\xi}$ (so that $y \in \cal
H^\xi$), $y$ be $C^\infty$-smooth in ${\mathbb R}^3$ outside the
sphere $\Gamma^{\xi_0}$, and
 \begin{align}\label{jump y (mathbb RR3^3)}
\lim \limits_{\varepsilon \to
0}\left[y\left((\xi_0+\varepsilon)\omega\right)-y\left((\xi_0-\varepsilon)\omega\right)\right]
\,=\,\alpha(\omega), \quad \omega \in S^2 .
 \end{align}
with a nonzero $\alpha \in C^\infty(S^2)$. Thus, we have
 $ \emptyset \not={\rm sing\, supp\,}y \subset \Gamma^{\xi_0}$ that yields
 \begin{equation*}
\emptyset \not= {\rm sing\, supp\,}v^y\,\subset\,C^{\xi_0}_1 \cup
 C^{\xi_0}_2,
 \end{equation*}
where $C^{\xi_0}_1:=\{x \in {\mathbb
R}^3\,|\,\,|x|=-t+\xi_0\},\,\, C^{\xi_0}_2:=\{x \in {\mathbb
R}^3\,|\,\,|x|=|t+\xi_0|\}$ are the characteristic cones (see Fig
1b and compare with (\ref{sing supp y (ball)}), (\ref{sing supp
v^y (ball)})). In particular, the cone $C^{\xi_0}_1$ supports a
jump of $v^y_t$ and $v^y_r$, whereas the geometrical optics
provides
 \begin{align}
\notag & \lim \limits_{\varepsilon \to
0}\left[v^y_r\left((\xi_0-t+\varepsilon)\omega,t\right)-v^y_r
\left((\xi_0-t-\varepsilon)\omega,t\right)\right]\,\\
 \label{jump d/dr v^y (mathbb RR3^3)}&=\,- \frac{\xi_0}{\xi_0-t} \alpha(\omega),
\qquad \omega \in S^2,\,\,t<0
 \end{align}
that is the same as (\ref{jump d/dr v^y (ball)}). So, the radial
derivative $v^y_r(\cdot,t)$ has a jump at $\Gamma^{\xi_0-t}$,
which moves (in the reversed time) to infinity. Up to the factor
$\frac{\xi_0}{\xi_0-t}$ of geometric nature, its shape reproduces
the shape of the initial jump (\ref{jump y (mathbb RR3^3)}) of the
Cauchy data.

By the use of technique \cite{BV3} (see Lemma 3) with regard to
(\ref{jump d/dr v^y (mathbb RR3^3)}), one can derive that the
limit passage (\ref{def O in mathbb RR3^3}) provides the relation
 \begin{align}\label{jump Oy (mathbb RR3^3)}
\lim \limits_{\varepsilon \to 0}\left[\left(O
y\right)(\xi_0+\varepsilon,\omega)-\left(Oy\right)(\xi_0-\varepsilon,\omega)\right]\,=\,-
\xi_0 \alpha(\omega), \quad \omega \in S^2,
 \end{align}
in perfect analogy to (\ref{jump O^Ty (ball)}). Thus, any jump of
the Cauchy data $y$ on a sphere propagates along the space-time
rays preserving its shape, reaches the infinity, and is observed
(as a jump of $Oy$) in the interval $\xi<\tau<\infty$ at the
moment $\tau=\xi_0$. Therefore, $(Oy)\big|_{[\xi,\infty)}\not=0$
and we can choose a control $f \in {\cal F}^\xi$ such that
 $$0\not=(Oy,f)_{\cal F}\overset{(\ref{O=W^*=
Radon})}=(y,Wf)_{\cal H}=\left(y,u^f(\cdot,0)\right)_{\cal
H}=\left(y,u^f(\cdot,0)\right)_{{\cal H}^\xi}
 $$
since $y \in {\cal H}^\xi$ and $u^f(\cdot,0)\in {\cal U}^\xi
\subset {\cal H}^\xi$. Hence, one can claim definitely that $y
\not \in {\cal D}^\xi$. In other words, no $h \in {\cal D}^\xi$
can have a jump on a sphere because it is visible at infinity. As
is easy to recognize, such a jump is also forbidden for any
derivative $\frac{\partial^k h}{\partial r^k},\,k\geqslant 1$.

\subsubsection*{Conjecture and counterexample}
Surely, the embedding ${\rm sing\,supp\,}y \subset \Gamma^{\xi_0}$
doesn't mean that the singularity of $y$ is necessarily a jump.
However, the above-considered example motivates to suggest that
elements of ${\cal D}^\xi$ are smooth at least with respect to
$|x|$ and, in particular, this embedding should be mutually
exclusive with $y \in {\cal D}^\xi$. One more pro-argument is that
${\cal D}^\xi$ is spanned on $C^\infty$-smooth polyharmonic
functions: see (\ref{D=sun Dl}), (\ref{polyharmonic}). So, we
conjectured that {\it any} spherical singularity is visible at
infinity.

The conjecture turns out to be wrong, the following is a
counterexample.
\smallskip

Take $\xi=1, \, \xi_0=2$; let $p(s):=3s-4s^3,\,\,\,s>0$. Choose a
(real) sequence $a_1, a_2, \dots$ provided
 \begin{equation}\label{sequence a}
\sum_{k\geqslant 1}a_k^2 <\infty,\,\,\,\,\sum_{k\geqslant 1}k^4
a_k^2 = \infty\,.
 \end{equation}
For every $l=6k+3\,\,\,(k=1,2, \dots)$, choose harmonics $Y_l\in
{\rm span}\{Y_l^m\}_{m=-l}^l$ such that
$(Y_l,Y_{l'})_{L_2(S^2)}=\delta_{l l'}$. Compose
 \begin{equation}\label{h}
h(x)=h(r,\omega):=
  \begin{cases}
0, & r<1\\
\sum \limits_{k \geqslant 1}a_k\,
\frac{1}{r}\left[p\left(\frac{1}{r}\right)\right]^{2k+1}Y_{6k+3}(\omega),
&r\geqslant 1\,.
  \end{cases}
 \end{equation}
The summands in (\ref{h}) rapidly decrees as $r\to \infty$, so
that $h$ is square-summable in ${\mathbb R}^3$ and $h \in {\cal
H}^1$ since ${\rm supp\,}h \subset \overline{\Omega^1}$. Moreover,
as is seen from (\ref{D=sun Dl}), one has $h \in {\cal D}^1$.

Let us show that $\emptyset \not={\rm sing\, supp\,}h \subset
\Gamma^2$ holds. For the polynomial $p$, one has
$p(\frac{1}{r})=\frac{3}{r}-\frac{4}{r^3}$ that implies
 \begin{equation}\label{p(1/r)}
\bigg|p\left(\frac{1}{r}\right)\!\bigg|<1\,\,\,\,{\rm for}\,\,r>1,
r\not=2\,;\quad p\left(\frac{1}{2}\right)=\max \limits_{r>1}
\bigg|p\left(\frac{1}{r}\right)\!\bigg|=1
 \end{equation}
(see Fig 2).
\begin{figure}[h]
 \begin{center}
 \epsfysize=8cm
 \epsfbox{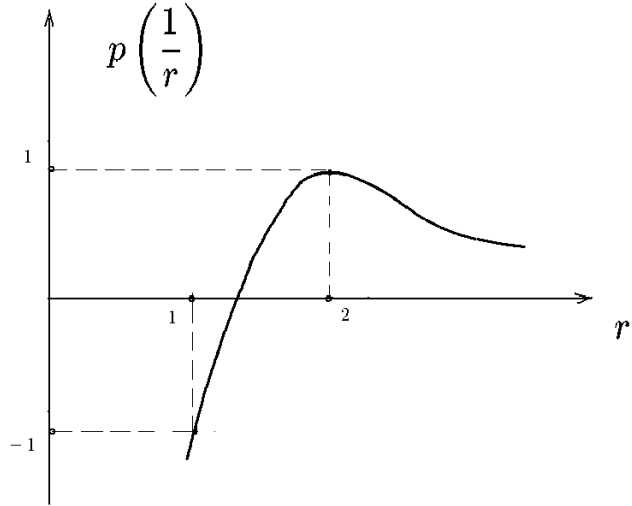}
 \end{center}
\caption{Polynomial $p$}
 \end{figure}
By the first property, for any fixed $r\not=2$, the series in
(\ref{h}) rapidly converges in $L_2(S^2)$ and
$h\left(r,\cdot\right) \in C^\infty(S^2)$ holds. The same can be
easily checked for the derivatives: $\frac{\partial^j h}{\partial
r^j}\big|_{r \not=2}\in C^\infty(S^2),\,\,j\geqslant 1$. As a
result, we have $h \in C^\infty(\Omega^1\backslash \Gamma^2)$.
Hence, ${\rm sing\,supp\,}h \subset \Gamma^2$.

Show that ${\rm sing\,supp\,}h$ is nonempty. Take $r=2$. By
(\ref{p(1/r)}), one has
 $$
h(2,\omega)\,=\,\frac{1}{2}\sum_{k \geqslant 1}a_k
Y_{6k+3}(\omega)\,, \quad \omega \in S^2\,,
 $$
so that $h(2,\cdot)\in L_2(S^2)$ by the choice of $a_k$ and
orthogonality of $Y_l$. In the mean time, applying the
Beltrami-Laplace operator, we have
 $$
\|\Delta_\omega h(2,\cdot)\|^2=\frac{1}{4}\sum_{k \geqslant
1}a_k^2\left[(6k+3)(6k+4)\right]^2\,\overset{(\ref{sequence
a})}=\,\infty\,.
 $$
Therefore, $h(2,\cdot)\not\in C^\infty(S^2)$, and the element
defined by (\ref{h}) does disprove the conjecture.
\smallskip

Note in addition that simple analysis provides for the derivatives
 $$
h_r(2,\cdot)=0, \quad \lim \limits_{r \to
2}\|h_{rr}(r,\cdot)\|_{L^2(S^2)}=\infty\,.
 $$
Also, we can make the counterexample even more expressive by
taking $a_k=1$. In this case, one has ${\rm sing\, supp\,}h
\subset \Gamma^2$ and $\lim \limits_{r \to
2}\|h(r,\cdot)\|_{L_2(S^2)}=\infty$.

\subsubsection*{Comments}
An open problem is to describe admissible singularities of
unobservable states belonging to $\cal D^\xi$. In particular, can
one characterize the spherical singularities {\it invisible at
infinity}? The case of a (visible) jump shows that one should use
more subtle distribution characteristics than $\rm sing\,supp$.
Probably, it is a wave front. For a jump (\ref{jump y (mathbb
RR3^3)}) with a positive $\alpha \in C^\infty(S^2)$, one has
$WF[y]=\{\gamma \times N_\gamma\backslash 0\,|\,\,\gamma \in
\Gamma^{\xi_0}\}$, where $N_\gamma = T_\gamma^*\Omega^\xi \ominus
T^*_\gamma \Gamma^{\xi_0}$ is the subspace conormal to
$\Gamma^{\xi_0}$ \cite{Horm}. However, to recognize, what is
$WF[h]$ of $h$ defined by (\ref{h}), seems to be rather difficult.
Does $WF[h]$ contain conormal directions? The question is also
open.

\bigskip

{\bf The authors:}

\bigskip
\noindent {MIKHAIL I.~BELISHEV;~~~SAINT-PETERSBURG DEPARTMENT OF
THE STEKLOV MATHEMATICAL INSTITUTE, RUSSIAN ACADEMY OF
SCIENCES;~~~belishev@pdmi.ras.ru}~(corresponding author)

\bigskip
\noindent {ALEKSEI F.~VAKULENKO;~~~SAINT-PETERSBURG DEPARTMENT OF
THE STEKLOV MATHEMATICAL INSTITUTE, RUSSIAN ACADEMY OF
SCIENCES;~~~vak@pdmi.ras.ru}

\bigskip
\noindent {\bf MSC:}~35Bxx,~35Lxx,~35P25,~47Axx

\bigskip
\noindent {\bf Key words:}~$3{\rm d}$ wave equation, control
problem,~reachable sets,~unobservable states
\end{document}